\documentclass[11pt]{article}
\usepackage{epsfig}
\usepackage{epstopdf}
\usepackage{amsmath}\usepackage{amssymb}
\usepackage{mathrsfs}
\usepackage{graphicx}

\newcommand {\be}{\begin{equation}}
\newcommand {\ee}{\end{equation}}
\newcommand {\ba}{\begin{eqnarray}}
\newcommand {\ea}{\end{eqnarray}}

\begin{document}

\def \a'{\alpha'}
\baselineskip 0.65 cm
\begin{flushright}
\ \today
\end{flushright}

\begin{center}{\large
{\bf Impact of Torsion Space-Time on $t\bar{t}$ observables at Hadron Colliders}} {\vskip 0.5 cm} {\bf Seyed Yaser Ayazi and Mojtaba Mohammadi Najafabadi}{\vskip 0.5 cm
} School of Particles and Accelerators, Institute for Research in
Fundamental Sciences (IPM), P.O. Box 19395-5531, Tehran, Iran
\end{center}

\begin{abstract}
Starting from the effective torsion space-time model, we study its effects on the top pair production cross section at hadron colliders. We also study the
effect of this model on top pair asymmetries at the Tevatron and the LHC. We find that torsion space-time can explain forward-backward asymmetry according to measured anomaly at Tevatron. We find an allowed region in the parameters space which can satisfy simultaneously all $t\bar{t}$ observables measured at Tevatron and LHC.

\end{abstract}

\section{Introduction}

Top quark physics is one of the most promising probe of beyond Standard model (SM) since top quark is the heaviest known particle and is copiously produced at hadron colliders. A very large number of top quarks are produced at the
LHC eventually more than $10^7$ $t\overline{t}$ pairs per year
\cite{Schilling:2012dx}. Moreover, due to the large mass of the top quark which is at the order of the electroweak symmetry breaking, its lifetime is very short. This feature causes that it decays before hadronzation and provide an opportunity to explore  precision test of the SM  and its various properties like charge, mass and spin.

Experimental results for the cross section of top pair at
Tevatron and LHC are well consistent with the SM prediction. Another important measurement for top quark production is top Forward-Backward  Asymmetry (FBA) at Tevatron. The CDF and D0
collaborations reported sizable measurement difference with SM prediction \cite{CDF,Afb} for FBA. Actually, this observation at
Fermilab Tevatron may already be a hint of new physics. In the SM, top
pair production can be produced via the $q\bar{q}$ annihilation
and $gg$-fusion. The interference between radiative corrections
involving gluon emission and also the interference of leading order diagram with box diagrams lead to FBA in the top
pair production \cite{Kuhn:1998kw}.

 Since the initial state of
proton-proton collisions at the LHC is symmetric, FBA is not observed but another asymmetry can be observed (Charge
Asymmetry) \cite{Kuhn:1998kw}.
Charge Asymmetry (CA) at the LHC is defined as the difference between events with
positive and negative absolute values of rapidities of top and
antitop quarks. The CMS collaboration has recently presented CA
measurement in $t\bar{t}$ production at the LHC for the center-of-mass energy of $7~\rm TeV$ and $1.09~\rm fb^{-1}$ of data. This
measurement is well consistent with the SM prediction.
Nevertheless, the measurement of CA at LHC can provide an
independent criterion to search for NP models which explain FBA at Tevatron. It seems difficult to explain the size and nature of FBA by any new extension of SM. Because experimental constraint from Tevatron and LHC on parameters space of these theories,  must be simultaneously satisfied and there seems to be a correlation between Tevatron FBA and LHC CA.

Despite the impressive success of SM, common point of view is that SM can not play the role of fundamental theory because at least it does not include quantum gravity. Therefore, desired fundamental theory is expected to provide the solution to the quantum gravity and maybe even explain the low energy observables.  Traces of such fundamental theories could be identified by some additional characteristics of the space-time, different from the SM fields. Space-time torsion is one of the candidates which can play this role. So far, the effect of torsion space-time on low-energy experiment have been studied in the literature \cite{Torsion effect}.

In order to explain  the measured FBA, many extensions of SM have been proposed. Some of these models propose unknown
heavy particles which can be exchanged in top pair production
process \cite{AguilarSaavedra:2012ma,Z',Ayazi:2012bb,Grinstein:2012pn,axigluon,twin Higgs model,KK gluon,color-triplet scalar,color-tsextet scalar,Khatibi:2012ug}.
In this paper, taking into account the possibility of torsion field effect on top pair production at hadron colliders and the consistency of this theory with all observables specially with FBA is studied.

The rest of this paper is organized as follows: In the next
section, we introduce effective approach to torsion and study effect of torsion field on top pair production at hadron colliders.  In
section~3, we summarize  observables which  we study at the LHC and
Tevatron and study the effects of torsion field exchange in production of $t\bar{t}$ at Tevatron and LHC.
We also calculate torsion effect on forward-backward top pair asymmetry at Tevatron and charge asymmetry at the LHC. This enables us to put the constraint on parameters space of torsion extension the SM by using the present experimental measurements. The conclusions are given
in section~4.

\section{The space-time Torion and its phenomenological aspects}
In this section, we briefly review notation of the gravity with torsion and quantum theory of matter fields in the external torsion field. In the following, we study the effects of torsion space-time on our observable specially top pair production at Tevatron and LHC.

In the theory with torsion field, $T^{\alpha}_{\beta\gamma}$ is defined as follows\cite{Shapiro:2001rz}:
\begin{eqnarray}
T^{\alpha}_{\beta\gamma}=\Gamma ^{\alpha}_{\beta\gamma}-\Gamma^{\alpha}_{\gamma\beta},
\label{exp}
\end{eqnarray}
where $\Gamma ^{\alpha}_{\beta\gamma}$ are the christopher symbols. In general case the torsion field can be presented in the irreducible components\cite{deBerredoPeixoto:1999vj}
\begin{eqnarray}
T_{\alpha\beta\gamma}=\frac{1}{3}T_{\beta}g_{\alpha\gamma}-T_{\gamma}g_{\alpha\beta}-\frac{1}{6}\varepsilon_{\alpha\beta\gamma\delta}S^{\delta}+q_{\alpha\beta\gamma},
\label{exp}
\end{eqnarray}
where the axial vector $S^{\delta}=\varepsilon^{\alpha\beta\gamma\delta}T_{\alpha\beta\gamma}$, $T_{\alpha}$ is a vector trace of torsion and $q_{\alpha\beta\gamma}$ is a tensor which satisfies the constraints $q^{\alpha}_{\beta\alpha}=0$ and $q_{\alpha\beta\gamma}\varepsilon^{\alpha\beta\gamma\delta}=0$.

In this paper, we calculate the contribution of exchange of torsion field to top pair production at hadron colliders. Let us now consider the interaction of Dirac spinor $\psi$ as external gravitational field with torsion which has the following form:
 \begin{eqnarray}
S_{f}=\int d^4x\sqrt{g}\{
i\overline{\psi}\gamma^{\mu}(\nabla_{\mu}-i\eta_1\gamma^{5}S_{\mu}+i\eta_2T_{\mu})\psi-m\overline{\psi}\psi\},
\label{exp}
\end{eqnarray}
where $\eta_1$ and $\eta_2$ are non-minimal parameters and $\nabla_{\mu}$ is Riemannian covariant derivative without torsion. For special case of minimal coupling this expression corresponds to the values $\eta_1=-1/8$ and $\eta_2=0$. It is shown in \cite{Buchbinder:1990ku} that for a fixed non-zero value of  $\eta_1$, this action is not renormalizable while the zero value for $\eta_2$ does not imply any difficulties. In this paper, we consider $\eta_1$ as an arbitrary parameter and take $\eta_2=0$. Therefore, the interaction between a Dirac field, with torsion is described by following action:
\begin{eqnarray}
S_{TS-matter}=i\int d^4x\sqrt{g}
\overline{\psi}_i(\gamma^{\alpha}\nabla_{\alpha}+i\eta_i\gamma^{5}\gamma^{\mu}S_{\mu}-i m_i)\psi_i\label{int1},
\end{eqnarray}
where $\eta_i$ is the non-minimal interaction parameter for corresponding spinor. Notice that this action corresponds to the complementary antisymmetric torsion  $T_{\alpha\beta\gamma}=-\frac{1}{6}\varepsilon_{\alpha\beta\gamma\delta}S^{\delta}$. In our case, we suppose that the metric is flat $g_{\mu\nu}=\eta_{\mu\nu}$. Unitarity and renormalizability conditions in effective low energy quantum field theory lead to free torsion action with this form:
\begin{eqnarray}
S_{TS-free}=\int d^4x(-\frac{1}{4}S_{\mu\nu}S^{\mu\nu}+\frac{1}{2}M^2_{TS}S_{\mu}S^{\mu}),
\label{exp}
\end{eqnarray}
where $M_{TS}$ is the mass of torsion and $S_{\mu\nu}=\partial_{\mu}S_{\nu}-\partial_{\nu}S_{\mu}$. As it has been seen in (Eq~.\ref{int1}), we can enter spinor-torsion interaction  to SM as interactions between fermions with a new vector field $S_{\mu}$. Therefore, the total action include ${\cal{L}}_{SM}$, ${\cal{L}}_{TS-matter}$ and ${\cal{L}}_{TS-free}$. In low energy limit, the total action leads to four fermions interaction term with the following form:
\begin{eqnarray}
{\cal{L}}_{int}=-\frac{\eta_a\eta_b}{M^2_{TS}}(\overline{\psi}_a\gamma_5\gamma^{\mu}\psi_a)(\overline{\psi}_b\gamma_5\gamma_{\mu}\psi_b),
\label{exp}
\end{eqnarray}
where $\eta_a$ and $\eta_b$ are dimensionless coupling constants. Therefore, the new four fermions interaction is characterized by dimensionless parameters $\eta_a$, $\eta_b$ and mass of the torsion field $M_{TS}$. In the following, we consider $\eta_a$, $\eta_b$ and $M_{TS}$ as arbitrary parameters and study the effect of exchange of torsion field on top pair production at the Tevatron and the LHC. The leading
order (LO) processes for the production of top pair at hadron colliders
include these two processes $q\overline{q}\rightarrow
t\overline{t}$ and $gg\rightarrow t\overline{t}$. We neglect interaction between non-abelian gauge fields (gluons) with torsion since interaction of purely antisymmetric torsion with gauge field does not save gauge invariant\cite{Shapiro:2001rz}.

We have calculated the
$q\overline{q}\rightarrow t\overline{t}$ including the above four fermion effects. When
$q\overline{q}$ is the initial state, the amplitude for top pair production is
given by:
\begin{eqnarray}
{\cal{|M|}}^2_{TS} & =\frac{\eta_t^2\eta_q^2s^2}{M^4_{TS}}[1+(1-\frac{4m_q^2}{s})\beta^2 \cos^2\theta-\frac{4m_q^2}{s}\beta^2],
\label{amp1}
\end{eqnarray}
where $s$ is partonic center-of-mass energy, $\beta=\sqrt{1-4m_t^2/s}$ , $m_{t}$ and $m_{q}$ are the top quark and $q$ quark masses. In
Eq. \ref{amp1}, $\theta$ is the production angle of the outgoing top in the center-of-mass system . Another contribution to this process arises from the interference between SM contribution and torsion effects.  Amplitude for this interference given by:
\begin{eqnarray}
\frac{16\eta_t\eta_q s g_s^2}{3M^2_{TS}}[1+(\sqrt{1-\frac{4m_q^2}{s}})\beta \cos\theta]^2,
\label{amp2}
\end{eqnarray}
where $g_s$ is the strong coupling constant. As we can see in the above amplitudes, the only quantity which appear in this approach is the ratio  $\frac{\eta_t\eta_q}{M^2_{TS}}$ and therefore for heavy torsion field the phenomenological consequences depend only on this single parameter.

\section{Observables and Numerical results}
In this section, we study the total cross section of top pair
production at the Tevatron and LHC and consider top pair forward
 backward asymmetry  and charge asymmetry as observables and
study the effect of torsion field on them.

Top pair production cross section at the Tevatron (by $\rm D0$ collaboration\cite{cross Tev}) and LHC7 (by CMS experiment \cite{cross LHC})  have been measured to be:
\begin{eqnarray}
\sigma_{\rm Tevatron}(pp\rightarrow t{\overline t} ) & =
7.56\pm0.83~[pb]~(\rm stat\oplus sys),\label{exp}
\end{eqnarray}
\begin{eqnarray}
\sigma_{\rm LHC}(pp\rightarrow t{\overline t} ) & =
165.8\pm13.3~[pb]~(\rm stat\oplus sys).\label{exp}
\end{eqnarray}
These measurements are in good  agreement with the SM predictions
\cite{SM Tev,SM LHC}. The total cross section of top pair production at hadron colliders
can be obtained by convoluting the partonic cross section with
the parton distribution functions (PDF) for the initial hadrons.
To calculate $\sigma(pp\rightarrow t\overline{t})$, we have used
the MSTW parton structure functions \cite{MSTW} and set the
center-of-mass energy to $7~ \rm TeV$ for the LHC and $1.96~ \rm TeV$ for the Teavatron. The total cross section
for production of $t\bar{t}$ is given by:
\begin{eqnarray}
\sigma(pp\rightarrow t{\overline {t}}) & = &\sum_{ab} \int
dx_1dx_2f_a(x_1,Q^2)f_b(x_2,Q^2) \widehat{\sigma}(ab\rightarrow
t{\overline {t}}), \
\end{eqnarray}
where $f_{a,b}(x_i,Q^2)$ are the parton structure functions of
proton. $x_1$ and $x_2$ are the parton momentum fractions and $Q$
is the factorization scale.

Here, we emphasis that for proton-antiproton collision FBA is
measured by using invariant difference of $t$ and $\bar{t}$ rapidities. The rapidity $y$ of the top quark is given by:
\begin{eqnarray}
\frac{1}{2}\ln(\frac{E+p_z}{E-p_z})
\end{eqnarray}
with $E$ being the total top quark energy and $p_z$ is the top quark momentum along beam axis. The definition of FBA is identical to asymmetry in the top production angle in the $t\bar{t}$ rest frame:
\begin{eqnarray}
A_{FB} &=&\frac{N_t(\cos\theta>0)-N_t(\cos\theta<0)}{N_t(\cos\theta>0)+N_t(\cos\theta<0)}.\
\end{eqnarray}
Prediction of SM for FBA is as small as a few percent which arises from the interference between the Born
amplitude for $q\bar{q}\rightarrow Q\bar{Q}$ and box diagrams and
the interference term between initial state radiation and final
state radiation \cite{Kuhn:1998kw}. At the Tevatron, since the
initial state is asymmetric (proton-antiproton collisions), the
top quark forward- backward asymmetry can be measured. Recent
measurements of FBA have been reported by  $\rm CDF$\cite{CDF} ($A_{FB}= 0.158\pm0.075$) and
$\rm D0$\cite{Afb} ($A_{FB}= 0.196\pm0.065$) collaborations which show $2\sigma$ deviation from the SM prediction.

 At the LHC, since initial state is symmetric (proton-proton collisions), FBA vanishes. However, charge asymmetry in $t\bar{t}$
production at LHC can be measured which reflects the top quark
rapidity distribution. The top quark charge asymmetry in top pair production is defined by \cite{Ac CMS}
\begin{eqnarray}
A_{C} & =
&\frac{N_t(\Delta(y^2)>0)-N_t(\Delta(y^2)<0)}{N_t(\Delta(y^2))>0+N_t(\Delta(y^2)<0)}\
\end{eqnarray}
where $\Delta(y^2)$ is,
\begin{eqnarray} \Delta(y^2) &
= &(y_t-y_{\bar{t}}).(y_t+y_{\bar{t}}).\
\end{eqnarray}

In proton-proton collisions at the LHC, $t\bar{t}$ has a boost  along the direction of the incoming quark, and therefore this leads to
a larger average rapidity for top quarks than anti-top quarks. As a result, CA in $t\bar{t}$ production is not zero.
The ATLAS and CMS collaboration have reported  the charge asymmetry measurements: $A_C
= -0.019 \pm0.036$ \cite{Ac Atlas}, $A_C =
-0.013\pm0.041$ ($A_C =
0.004\pm0.014$) \cite{Ac CMS} , and the SM prediction is $A_C =
0.0115$ \cite{Ac SM}. Note that measurement of charge
asymmetry at the LHC is in agreement with SM prediction while measurements of the FBA
show a deviation from the SM expectations. It means
that any new physics which explains the $t\bar{t}$
forward-backward asymmetry must satisfy $A_C$ measurements
consistent with the SM predictions. We are also interested in the differential cross section as a
function of invariant mass
$M_{t\overline{t}}=\sqrt{(p_t+p_{\overline{t}})^2}$, where $p_t$
and $p_{\overline{t}}$ are the four-momenta of top and anti-top,
respectively. This quantity is defined by
\begin{eqnarray}
\frac{d\sigma(pp\rightarrow
t\overline{t})}{dM_{t\overline{t}}}&=&\sum_{ab}\int_{\frac{M^2_{t\overline{t}}}{E^2_{CMS}}}^1dx_1[f_a(x_1,Q^2)
f_b(\frac{M^2_{t\overline{t}}}{x_1E^2_{CMS}},Q^2)
\\ \nonumber &\times& \frac{2M^2_{t\overline{t}}}{x_1E^2_{CMS}}\widehat{\sigma}(ab\rightarrow
t{\overline t})],
\end{eqnarray}

As it is mentioned, physical observables related to torsion field depend on the the mass of torsion $M_{TS}$ and coupling constant between torsion and fermion field $\eta_{\psi}$. In \cite{Chatrchyan:2011ns}, a search for narrow resonances has been performed by CMS experiment at LHC and any resonance below $1~\rm TeV$ has been excluded. In this paper, we consider mass of torsion field in energy scale more than $1~\rm TeV$. For the sake of simplicity, we choose identical couplings for interaction between torsion field and fermions. This assumption enables us to put the limit in two dimensional parameters space ($M_{TS}-\eta$) using the present experimental measurements.

To perform our analysis, we have set $m_t=172.5~\rm GeV$ and
fixed renormalization and factorization scale $\mu_R=\mu_F=m_t$.
For including higher order QCD effects, we have normalized the cross section  to the ratio of measured experimental cross section and
the leading order SM cross section.

\begin{figure}
\begin{center}
\centerline{\epsfig{figure=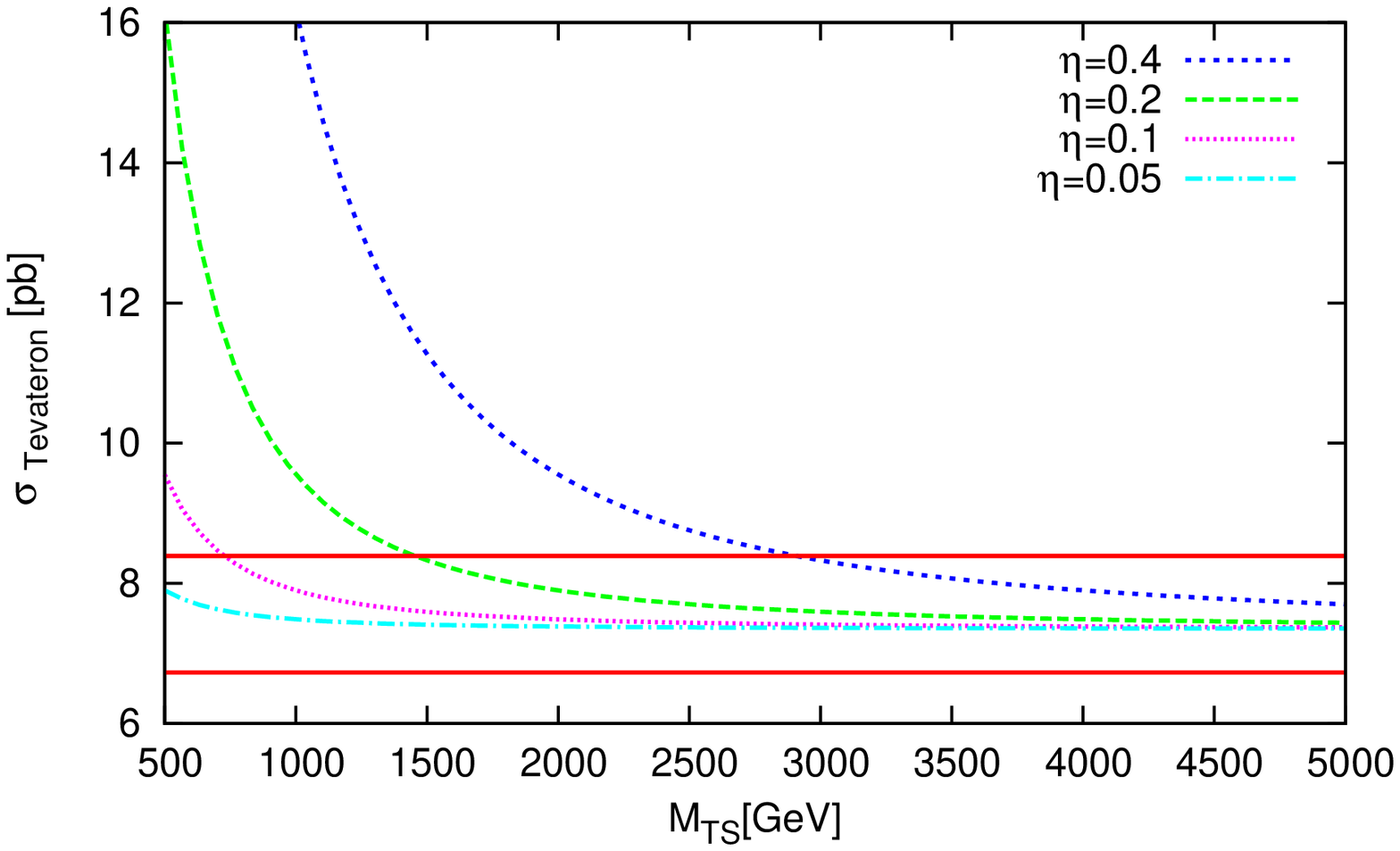,width=7.5cm}\hspace{5mm}\epsfig{figure=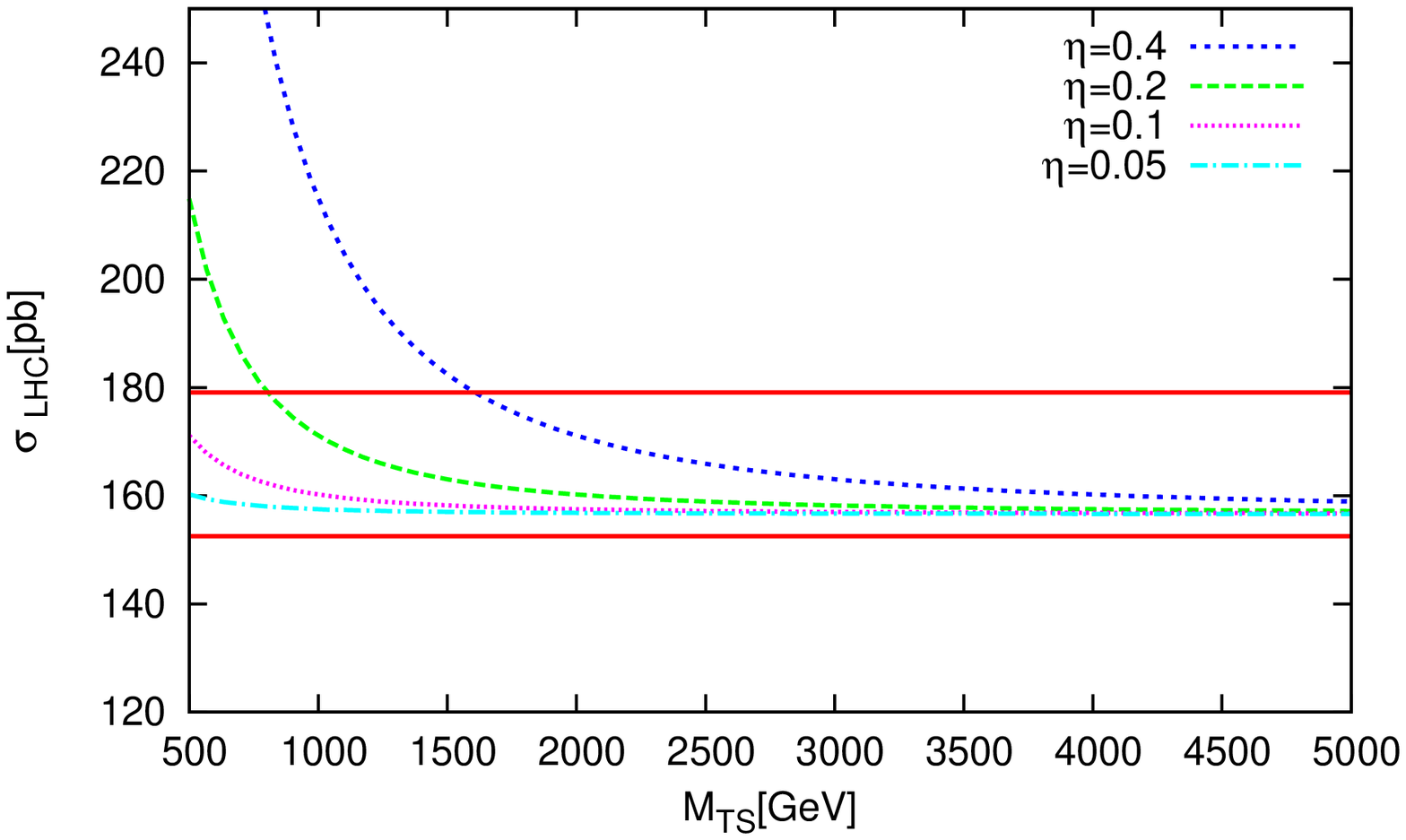,width=7.5cm}}
\centerline{\vspace{2.cm}\hspace{0.5cm}(a)\hspace{7cm}(b)}
\centerline{\vspace{-3.5cm}}
\end{center}
\caption{The top pair production cross section as a function of
the torsion mass at Tevatron (a) and LHC (b). The horizontal red
lines show allowed range of experimental measurements for the top
pair total cross section. } \label{cross}
\end{figure}

In Fig. \ref{cross}, we have
displayed  the total cross section of top-antitop production at
the Tevatron and LHC as a function of torsion field mass. The curves with different
colors and lines show various values of constant coupling $\eta$.

As it is shown in Eq.~\ref{amp1}, amplitude for torsion exchange depend on $\rm cos^2\theta$ which is symmetric after integrating over $\rm cos~\theta$. However Eq.~\ref{amp2} shows that interference term depend on $\rm cos~\theta$. As a result,  torsion exchange can contribute to FBA via the interference term with the SM contribution.

\begin{figure}
\begin{center}
\centerline{\epsfig{figure=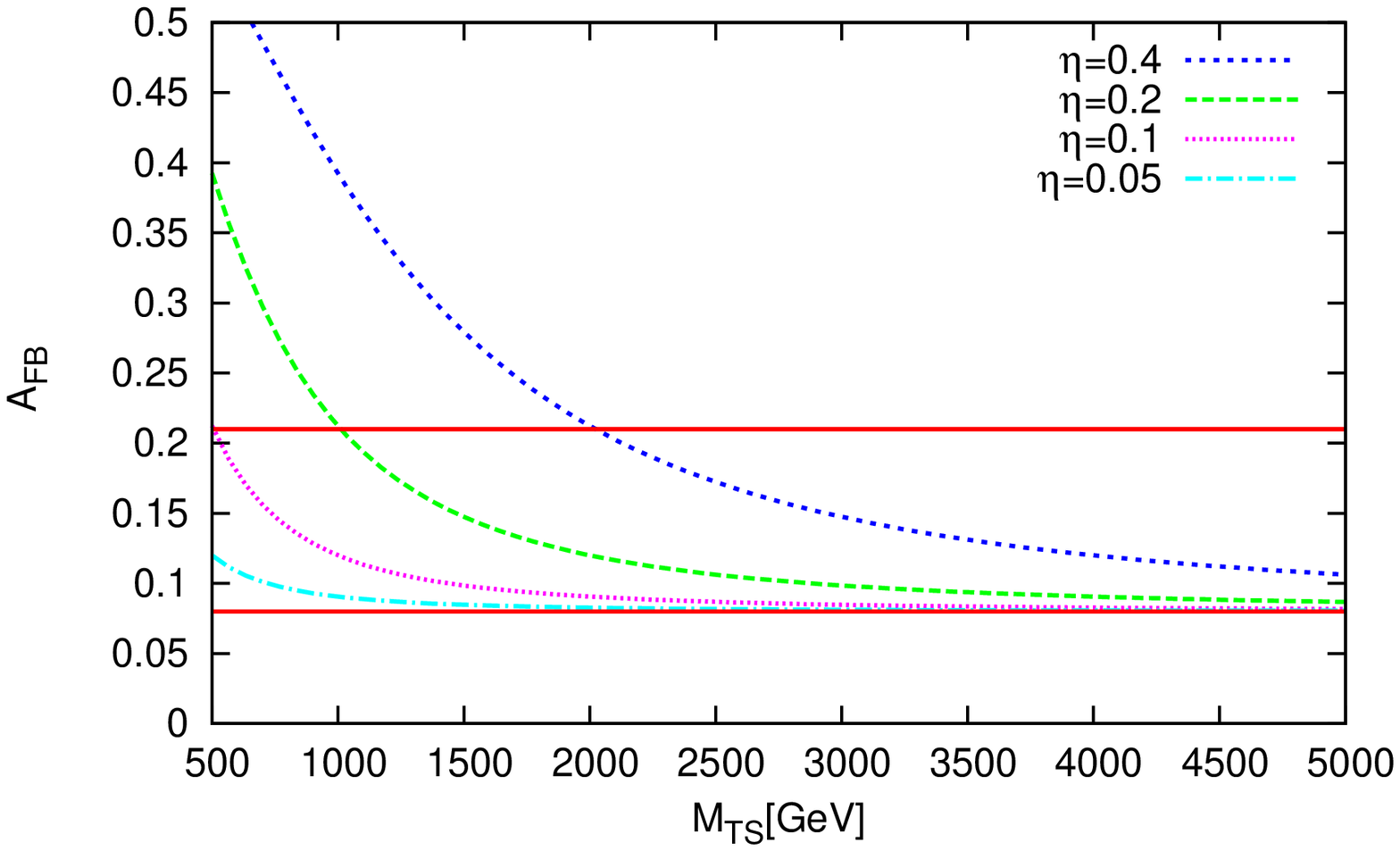,width=7.5cm}\hspace{5mm}\epsfig{figure=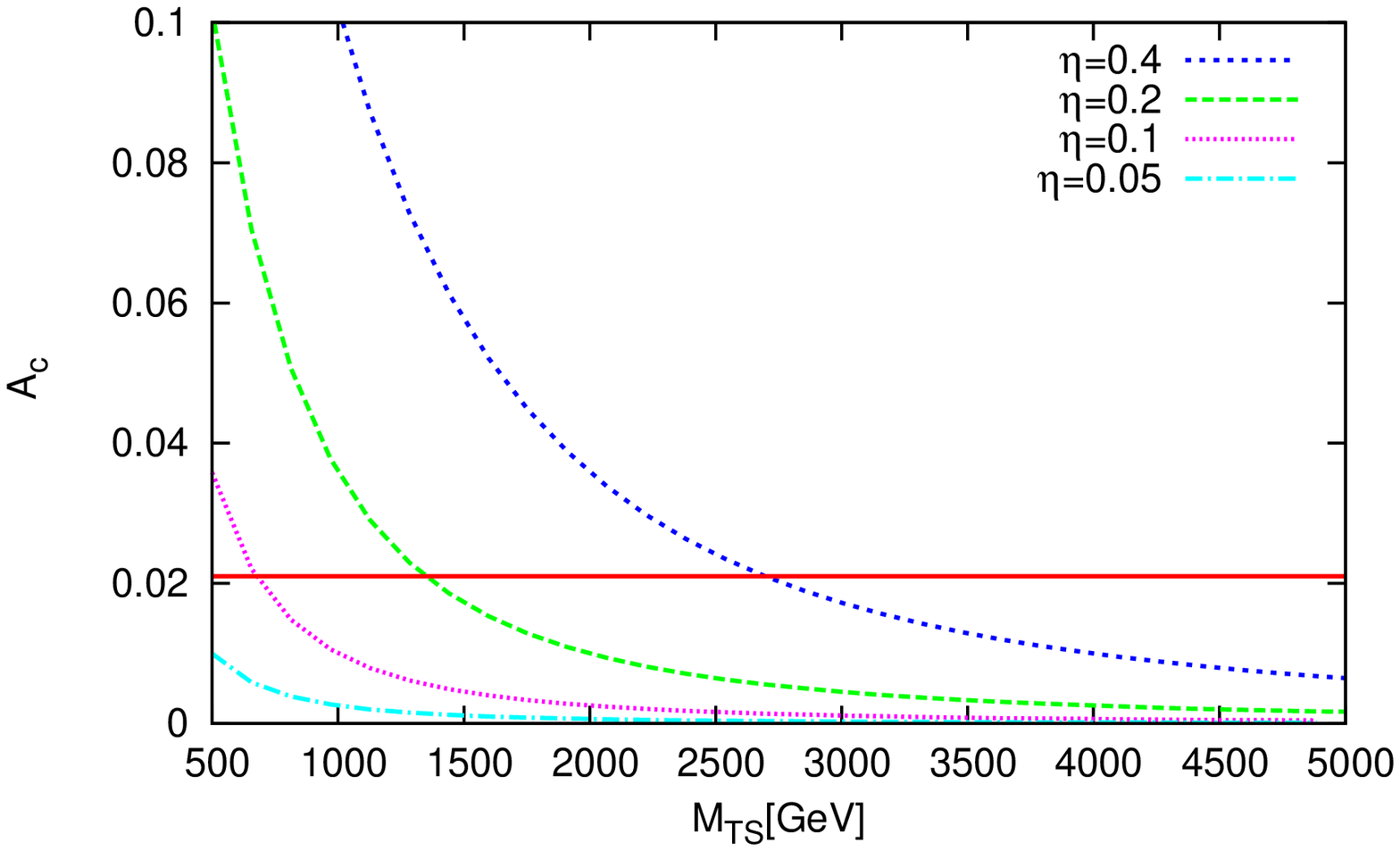,width=7.5cm}}
\centerline{\vspace{1.cm}\hspace{0.5cm}(a)\hspace{7cm}(b)}
\centerline{\vspace{-2.5cm}}
\end{center}
\caption{ The Top pair asymmetries as a function of  torsion mass.
a) Forward-Backward asymmetry at Tevatron. b) Charge asymmetry at
LHC. The horizontal red lines show the allowed ranges of
experimental measurements for asymmetries.} \label{Asy}
\end{figure}

Fig.~\ref{Asy}-a(b) depicts $A_{FB}$($A_C$) at Tevatron (LHC).
Different values for couplings have been considered. As it can be seen, for instance the
allowed values for $\eta=0.2$ in $A_{FB}$($A_C$) curves are
satisfied for $M_{TS}>1100~\rm GeV$ ($M_{TS}>1300~\rm GeV$).
This means that for a given value $\eta=0.2$, there are allowed regions in parameters space which are consistent with top asymmetries measurements.

\begin{figure}
\begin{center}
\centerline{\hspace{-1cm}\epsfig{figure=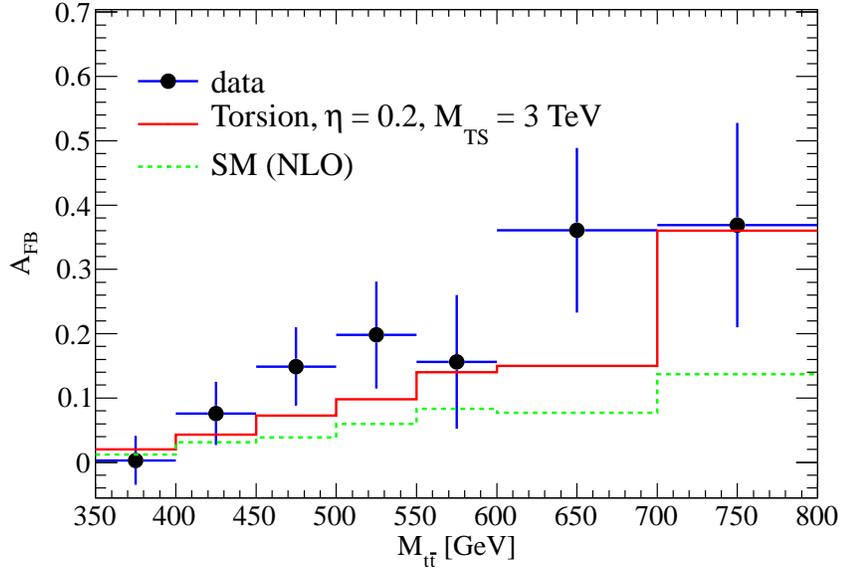,width=12cm}}
\centerline{\vspace{-1.5cm}}
\end{center}
\caption{Forward-backward asymmetry as function of invariant mass $M_{t\bar{t}}$ for torsion model compared to the experimental measurement at Tevatron and SM (NLO) expectation.} \label{histogram}
\end{figure}

The dependence of $A_{FB}$ on invariant mass of $t\bar{t}$ are studied in Fig.~\ref{histogram}. In this figure, we display a histogram which depicts forward-backward asymmetry as a function of invariant mass $t\bar{t}$ with seven bins. The green distribution shows SM expectation at to next leading order and red distribution shows measured data at Tevatron \cite{Aaltonen:2012it}. The uncertainties on the data contain statistical and systematics. The blue graph shows torsion model prediction in each bin. To draw this histogram, we set $\eta=0.2$ and $M_{TS}=3~\rm TeV$. This histogram  compare the experimental measurement at Tevatron and SM (NLO) with torsion model prediction. It is remarkable that the SM predictions in each bin are far from measured data at Tevatron and torsion prediction is approximately correspond with data.

\begin{figure}
\begin{center}
\centerline{\hspace{3.5cm}\epsfig{figure=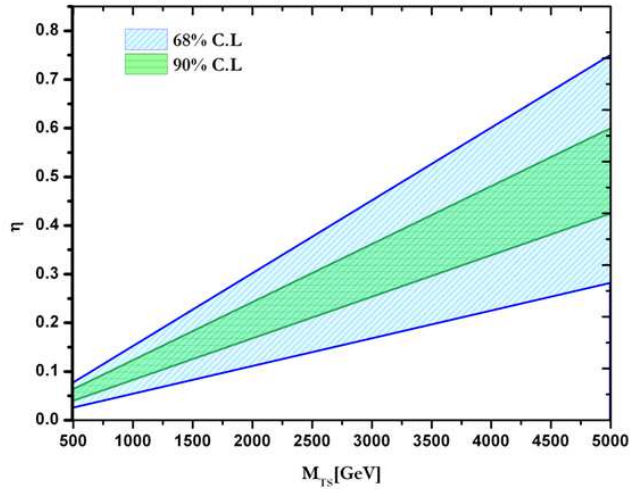,width=12cm}}
\centerline{\vspace{-3.5cm}}
\end{center}
\caption{Shaded areas depict ranges of parameters space in torsion mass $M_{TS}$  and coupling constant plane for which are consistent with $68\% ~\rm C.L$ and $90\% ~\rm C.L$.} \label{scaterX2}
\end{figure}

We perform a $\chi^2$-fit for forward-backward asymmetry as a function of invariant mass $t\bar{t}$ system ($M_{t\bar{t}}$). This quantity has been defined as:
\begin{eqnarray}
\chi^2=\sum_{i}(\frac{(A_{FB})_i-(A^{exp}_{FB})_i}{(\sigma^{exp})_i})^2,\label{Xai}
\end{eqnarray}
where summation is over all bins of invariant mass. The size of bins have been chosen according to bins in Fig.~\ref{histogram}. In Fig.~\ref{scaterX2}, we show range of parameters space in torsion mass of $M_{TS}$  and coupling constant $\eta$ plane which are consistent with $68\% ~\rm C.L$ and $90\% ~\rm C.L$. The blue area shows regions with $1~\sigma$ ($68\% \rm C.L$)  and green area depicts regions  with $1.7~\sigma$ ($90\% \rm C.L$) deviation from measured data in global analyze.

\begin{figure}
\begin{center}
\centerline{\hspace{-2cm}\epsfig{figure=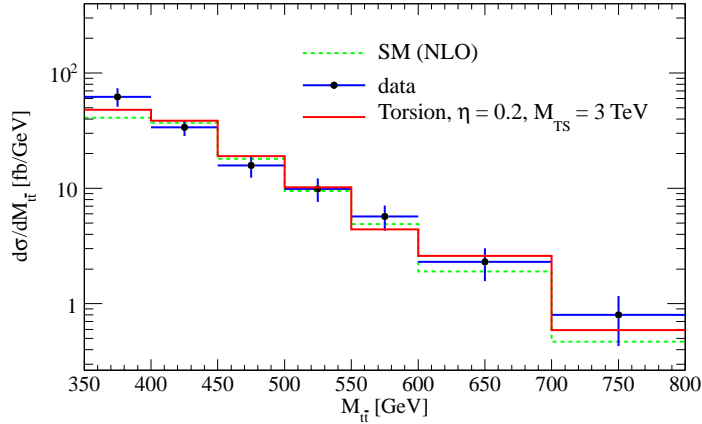,width=10cm}}
\centerline{\vspace{-1.5cm}}
\end{center}
\caption{Differential cross section of top-antitop production at
the Tevatron as a function of invariant mass $M_{t\overline{t}}$ for torsion model compared to the experimental measurement at Tevatron and SM (NLO) expectation.} \label{dMtt}
\end{figure}

Another observable which can help us to study torsion space-time effects on top pair production is differential cross section. Fig. \ref{dMtt} depicts differential cross section of $t\bar{t}$ production at Tevatron as a function of invariant mass $t\bar{t}$ system. The green distribution shows SM expectation at to next leading order and red distribution shows measured data at Tevatron \cite{Aaltonen:2009iz}. The uncertainties on the data contain statistical and systematics and luminosity. The blue graph shows torsion model prediction in each bin. To draw this histogram, we set $\eta=0.2$ and $M_{TS}=3~\rm TeV$. This histogram  compares the experimental measurement at Tevatron and SM (NLO) with torsion model prediction. As it is shown, the consistency  between torsion model prediction with measurement is reasonable.

\begin{figure}
\begin{center}
\centerline{\hspace{3.5cm}\epsfig{figure=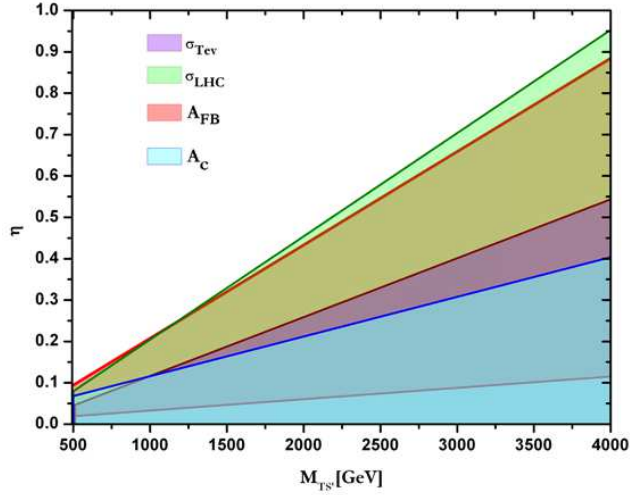,width=12cm}}
\centerline{\vspace{-3.5cm}}
\end{center}
\caption{Shaded areas depict ranges of parameters space in torsion mass $M_{TS}$  and coupling constant plane for which are consistent with
experimental measurements of observables: $\sigma_{\rm LHC}$,
$\sigma_{\rm Tev}$, $A_{\rm FB}$ and $A_{\rm C}$. } \label{scater}
\end{figure}

As it was mentioned in explanation of Fig.~\ref{Asy}, there are allowed regions in parameters space which are consistent with top asymmetries measurements. However, this result may conflict with allowed regions for cross section measurement at Tevatron and LHC. To better study of all parameters space which
 can simultaneously satisfy experimental constraints on $\sigma_{\rm
Tevatron}$, $\sigma_{\rm LHC}$, $A_{FB}$ and $A_{C}$, we display
Figs.~\ref{scater}. In Fig.~\ref{scater}, the shaded areas depict ranges of parameters space in torsion mass $M_{TS}$  and coupling constant plane for which prediction of effect of torsion space-time on observables $\sigma_{\rm LHC}$,
$\sigma_{\rm Tev}$, $A_{\rm FB}$ and $A_{\rm C}$ are consistent with experimental measurements. It is notable that the allowed regions of top pair production cross section at the LHC and Tevatron, overlap with allowed regions of $A_C$ and
$A_{FB}$. Also there is an overlapping region between the measured
$A_C$ at LHC and measured $A_{FB}$ at Tevatron which increases at high torsion mass $M_{TS}$. It is remarkable that we find regions in torsion parameters space which all experimental measurement are simultaneously satisfied. This means torsion model can explain measured forward-backward anomaly at Tevatron and top pair cross section measurement and charge asymmetry measurement at LHC can not exclude this model within the present uncertainty.

\section{Concluding remarks}
In this paper, we have studied the effects of torsion space-time on top pair asymmetries and the cross section productions at the
Tevatron and the LHC. We studied the dependence of $A_{FB}$ on invariant mass of $t\bar{t}$. It is shown that prediction of torsion model is more consistent with measured data at Tevatron than SM prediction for $A_{FB}$ in each bins of $M_{t\bar{t}}$. We performed global $\chi^2$-fit on forward-backward asymmetry as a function of invariant mass $M_{t\bar{t}}$ and showed regions with  $1~\sigma$ and $1.7~\sigma$ deviation from measured data in effective torsion model.  We also investigate the differential cross section of $t\bar{t}$ production as independent observable  and showed the consistency of this observable for torsion model with data.

Lastly, We have found overlapping regions in parameters space of torsion space-time which all experimental measurement ($\sigma_{\rm LHC}$,
$\sigma_{\rm Tev}$, $A_{\rm FB}$ and $A_{\rm C}$) are simultaneously satisfied. This means torsion model can explain measured forward-backward anomaly at Tevatron and top pair cross section measurement and charge asymmetry measurement at LHC can not exclude this model.

\section{Acknowledgement}
S. Y. A would like to acknowledgement CERN theory division for their hospitality.

\end{document}